\documentclass[12pt]{iopart}
\usepackage{graphicx}
\eqnobysec

\newcommand{\be}{\[}
\newcommand{\bea}{\begin{eqnarray*}}
\newcommand{\beq}{\begin{equation}}
\newcommand{\beqa}{\begin{eqnarray}}
\newcommand{\ee}{\]}
\newcommand{\eea}{\end{eqnarray*}}
\newcommand{\eeq}{\end{equation}}
\newcommand{\eeqa}{\end{eqnarray}}
\renewcommand{\a}{{\scriptscriptstyle\rm A}}

\renewcommand{\d}{{\rm d}}
\newcommand{\eps}{\varepsilon}
\newcommand{\eq}{{\rm st}}

\newcommand{\frad}[2]{\displaystyle{\displaystyle#1\over\displaystyle#2}}
\newcommand{\gl}{{\rm gl}}
\renewcommand{\i}{{\rm i}}
\newcommand{\infy}{{(\infty)}}
\newcommand{\la}{\lambda}
\newcommand{\loc}{{\rm loc}}
\newcommand{\ma}{{\scriptscriptstyle\rm MA}}
\renewcommand{\max}{{\rm max}}
\newcommand{\mean}[1]{\langle#1\rangle}
\newcommand{\mimi}{{\rm mean}}
\renewcommand{\min}{{\rm min}}
\newcommand{\noneq}{{\rm non-st}}

\newcommand{\s}{{\sigma}}
\newcommand{\sg}{{\rm sg}}

\newcommand{\C}{{\rm C}}
\newcommand{\D}{\Delta}
\newcommand{\E}{{\cal E}}
\newcommand{\F}{{\cal F}}
\newcommand{\Int}{\mathop{\rm Int}\nolimits}
\renewcommand{\L}{{\rm L}}
\newcommand{\M}{{\cal M}}
\renewcommand{\P}{{\cal P}}
\newcommand{\R}{{\rm R}}

\begin{document}
\title{Dynamics of the condensate in zero-range processes}
\author{C Godr\`{e}che\dag\ and J M Luck\ddag}

\address{\dag\ Service de Physique de l'\'Etat Condens\'e,
CEA Saclay, 91191 Gif-sur-Yvette cedex, France}

\address{\ddag\ Service de Physique Th\'eorique\footnote{URA 2306 of CNRS},
CEA Saclay, 91191 Gif-sur-Yvette cedex, France}

\begin{abstract}
For stochastic processes leading to condensation, the condensate,
once it is formed, performs an ergodic stationary-state motion over the system.
We analyse this motion, and especially its characteristic time,
for zero-range processes.
The characteristic time is found to
grow with the system size much faster than the diffusive timescale, but not
exponentially fast.
This holds both in the mean-field geometry and on finite-dimensional lattices.
In the generic situation where the critical mass distribution follows a power
law, the characteristic time grows as a power of the system size.
\end{abstract}


\eads{\mailto{godreche@dsm-mail.saclay.cea.fr},
\mailto{luck@dsm-mail.saclay.cea.fr}}


\maketitle

\section{Introduction}

In recent years many studies have been devoted to
nonequilibrium statistical-mechanical models yielding condensation,
such as urn models~\cite{zeta1,zeta2,zeta3,barc},
zero-range processes (ZRP)~\cite{loan,evans1,cg,gross,evans2}, generalised mass
transport models~\cite{maj},
or driven diffusive systems~\cite{wis1,wis2,pk,glm}.
In all these models except for the latter, the condensate manifests itself by
the macroscopic occupation of a single site
of a thermodynamically large system by a finite fraction of the whole mass.
For driven diffusive systems, the condensate manifests itself as a domain
of macroscopic size.

For a large but finite system at stationarity, the condensate is expected
to perform some ergodic motion over the sample.
The aim of this work is to investigate in detail the nature
of this motion in the stationary state, on the example of a class of ZRP.
We demonstrate that this motion is fully non-local, and we determine
the corresponding characteristic time $\tau$.
In the situation of most current interest where the ZRP has a power-law
critical mass distribution,
we find that $\tau$ scales as a power of the system size, in any dimension.
Finally, our predictions are generalised to other types
of critical mass distributions.
Whatever this distribution, $\tau$ never scales exponentially with the system
size.
Little attention had been devoted so far to the stationary-state
motion of the condensate~\cite{gross,maj}.
A critical reading of these references is given at the end of the present work.

\section{Definition of the model}

A ZRP is a stochastic process, defined as follows~\cite{spitz,andj,evans2}.
Consider a finite connected graph, made of~$M$ sites.
Particles hop from site to site with rates
which only depend on the occupation of the departure site.
Denote by $N_i$ the instantaneous occupation of site~$i$ ($i=1,\dots,M$).
The total number of particles $N_1+\cdots+N_M=N$ is conserved by the dynamics.

An elementary step of the dynamics consists in choosing a departure site~$d$
and an arrival site~$a$ connected to site~$d$,
and in transferring one of the particles present on site~$d$ to site~$a$.
This process takes place with rate $u_k$ per unit time, where $k=N_d\ne0$.
In the mean-field geometry, all sites are connected,
i.e., sites~$d$ and $a$ are chosen independently at random.
On finite-dimensional lattices, site~$a$ is chosen among the first neighbours
of site~$d$, possibly in an asymmetric way.
In one dimension, the most general dynamics corresponds to choosing site~$a$
to be the right neighbour of site~$d$ with probability $p$,
or its left neighbour with the complementary probability $q=1-p$.
In the following we only consider the one-dimensional symmetric dynamics,
corresponding to $p=1/2$, and the fully asymmetric one,
corresponding to $p=1$, both with periodic boundary conditions.

The fundamental property of ZRP is that
the stationary distribution of particles is factorised,
and explicitly known in terms of the rates $u_k$~\cite{spitz,evans2}.
The probability of a given configuration of the system $\{N_1,\dots,N_M\}$
is equal to a product of weights associated with the occupation of each site
(see Appendix~A for more details):
\beq
P(\{N_i\})=\frac{1}{Z_{M,N}}\prod_{i=1}^M p_{N_i},
\label{eq:pn}
\eeq
where
\beq
p_0=1,\qquad p_k=\frac{1}{u_1\dots u_k}.
\label{pk}
\eeq
The normalisation factor $Z_{M,N}$ plays the role of a partition function.
The stationary occupation probabilities of any given site read
\beq
f_{k,\eq}=\frac{p_k\,Z_{M-1,N-k}}{Z_{M,N}}.
\label{fkeq}
\eeq

In the present work, unless otherwise stated, we consider the
rate~\cite{evans1,wis1,cg,gross}
\beq
u_k=1+\frac{b}{k},
\label{rates}
\eeq
where $b$ is a given parameter, which plays the role of inverse temperature.
With this choice of rate,
in the thermodynamic limit ($M\to\infty$ at fixed density $N/M=\rho$),
the system has a continuous phase transition at the critical density
\be
\rho_c=\frac{1}{b-2},
\ee
whenever $b>2$, i.e., at low enough temperature.
The critical density separates a fluid phase from a condensed phase.
In the fluid phase $(\rho<\rho_c)$,
the occupation probabilities $f_{k,\eq}$ fall off exponentially.
At the critical density $(\rho=\rho_c)$, they fall off as a power law:
\beq
f_{k,\eq}\approx\frac{(b-1)\Gamma(b)}{k^b}
\label{fkc}
\eeq
(see~(\ref{afkc})).
In the condensed phase $(\rho>\rho_c)$, for a large and finite system,
the particles form a uniform critical background
and a macroscopic condensate, consisting (on average)~of
$\D$ excess particles with respect to the critical state, where
\beq
\D=N-M\rho_c=M(\rho-\rho_c).
\label{deldef}
\eeq
The condensate appears as a hump in the stationary distribution
$f_{k,\eq}$ (see Figure~\ref{ffk} below).

The case of the rate~\cite{wis1,wis2}
\be
u_k=1+\frac{a}{k^\s},
\ee
where $\s$ is an arbitrary exponent,
is discussed at the end of the present work.

\section{Heuristic approach}

For a large and finite system in the stationary state, as time passes,
the condensate keeps moving across the system:
it spends long lengths of time on a given site,
before suddenly disappearing and reappearing on another site.
The typical value of these lengths of time defines the characteristic time
$\tau$ of the dynamics of the condensate.
The aim of this work is to characterise
how $\tau$ scales with the system size $M$.

An intuitive understanding of the phenomenon is easily
gained by performing simple Monte-Carlo simulations.
Such simulations done
in the three geometries introduced above: mean-field (MF),
one-dimensional asymmetric (1DAS) ($p=1$),
and one-dimensional symmetric (1DS) ($p=q=1/2$), lead to a common picture.

\begin{figure}[htb]
\begin{center}
\includegraphics[angle=90,width=.6\linewidth]{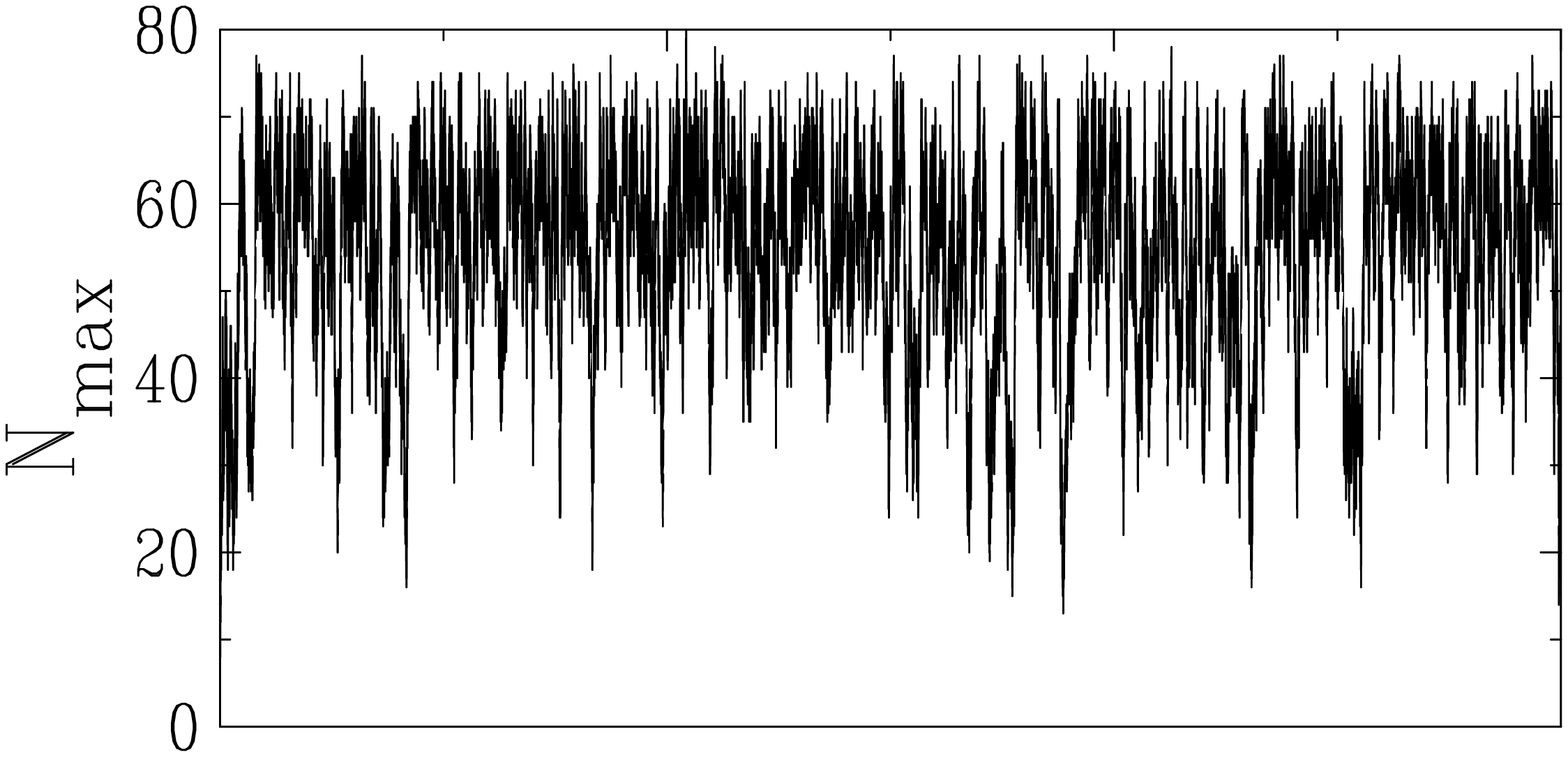}
\vskip -9truemm
\includegraphics[angle=90,width=.6\linewidth]{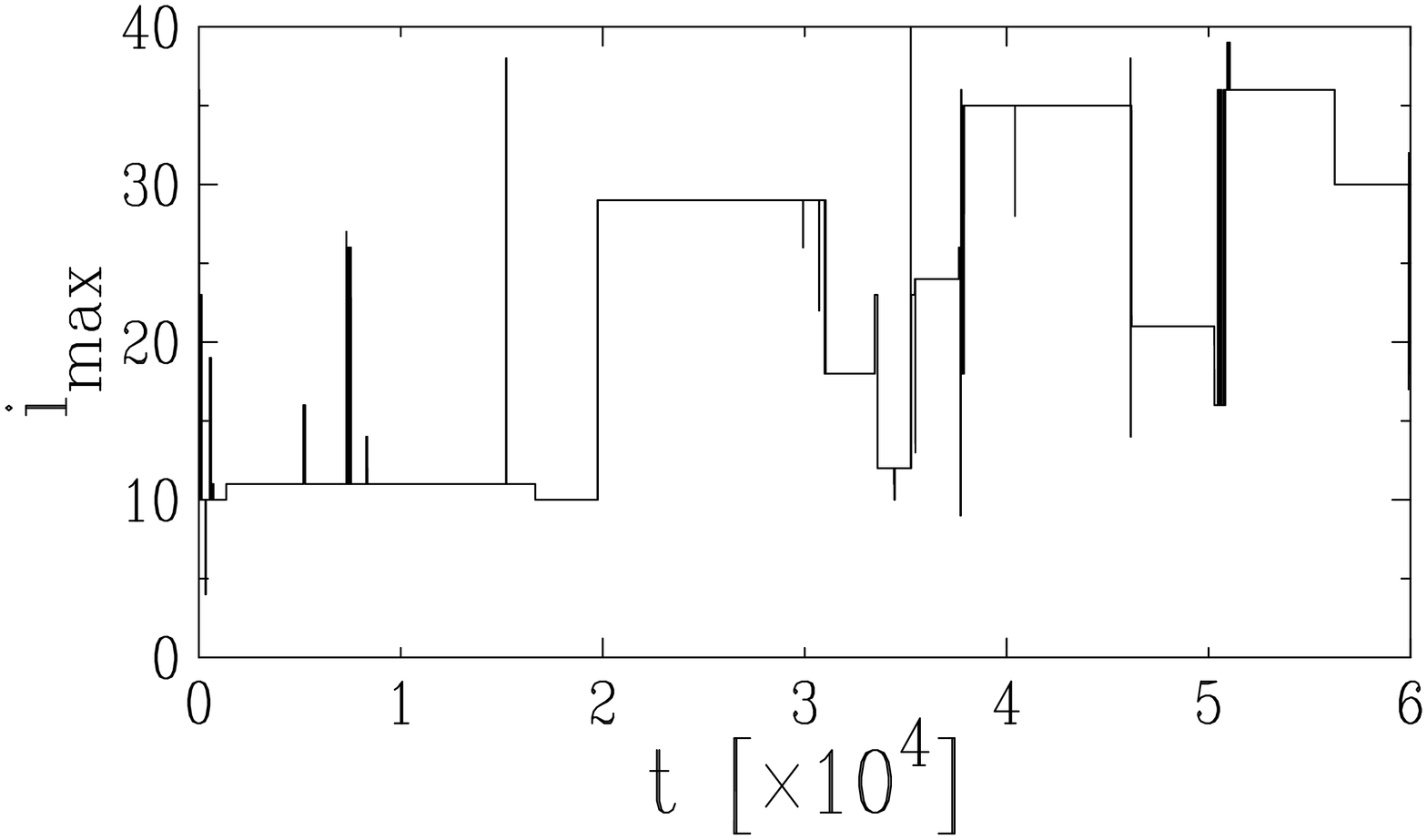}
\caption{\small
Dynamics of the condensate (1DS geometry with $b=4$, $M=40$, $N=80$).
Upper panel: instantaneous number of particles
$N_\max(t)$ on the most populated site.
Lower panel: location $i_\max(t)$ of that site.}
\label{fhist}
\end{center}
\end{figure}

The condensate is immobile for rather long lapses of time;
it then performs sudden random non-local jumps all over the system,
at Poissonian times whose characteristic scale
grows rapidly with the system size $M$.
Figure~\ref{fhist} illustrates this process for the 1DS case,
for a system of size $M=40$, with $N=80$ particles, i.e., $\rho=2$,
and $b=4$, hence $\rho_c=1/2$.
The upper panel shows the track of the instantaneous number of particles
$N_\max(t)$ on the most populated site.
The signal for $N_\max(t)$ fluctuates around $\Delta=60$,
the mean size of the condensate (see~(\ref{deldef})).
The lower panel shows the label $i_\max(t)$ of that site,
i.e., the location of the condensate.
The non-local character of the motion of the condensate is clearly visible,
whereas the longest lapses of time where the condensate stays still
give a heuristic measure of the characteristic time $\tau$.

These features will be turned to quantitative predictions in the following.
We first analyse in detail the case of two sites $(M=2)$ (Section~\ref{sectwo}).
We then introduce an approximate model for the dynamics of the ZRP, for $M$
arbitrary, based on a Markovian Ansatz (Section~\ref{secma}).
The relaxation time $\tau_\ma$ within the Markovian Ansatz
will be shown to grow asymptotically according to~(\ref{zrpt1}).
Its dependence on $M$ and $\rho$ is the following:
\be
\tau_\ma\sim(\rho-\rho_c)^{b+1}M^b\sim\frac{\D^{b+1}}{M}.
\ee
We finally present the results of Monte-Carlo simulations
for the temporal correlations of local and global density fluctuations
in the stationary state of the ZRP (Section~\ref{secsim}).
The associated respective relaxation times
$\tau_\loc$ and $\tau_\gl$ will be shown
to scale as $\tau_\ma$ in the MF and 1DAS geometries,
and as $M\tau_\ma\sim\D^{b+1}$ in the 1DS geometry, up to finite prefactors.

For clarity we have regrouped in two appendices the relevant tools for the
derivation of the results of sections 4 and 5.
Appendix A is concerned with the static properties of ZRP,
Appendix B with the analysis of the characteristic times appearing
in the biased motion of a random walker on an interval.

\section{The case of two sites}
\label{sectwo}

The case of two sites ($M=2$, $N$ arbitrary) is exemplary because it shares
a number of features with the general case.
In the stationary state, for $N$ large, the system flips between two states
where most of the particles are either on one site, or on the other one.
In this section we determine the flipping time of the system, that is
the characteristic time for the system to flip between these two states.

The dynamics of the two-site system is encoded in the
master equation for the occupation probability of site~1,
defined as $f_k(t)=\P(N_1(t)=k)$.
The temporal evolution of this quantity is given by the equations
\beqa\label{2sites}
\frac{\d f_k(t)}{\d t} &=&\mu_{k+1}\,f_{k+1}+\la_{k-1}\,f_{k-1}-(\mu
_k+\la_k)f_k\qquad(1\le k\le N-1),\nonumber\\
\frac{\d f_0(t)}{\d t} &=&\mu_1\,f_1-\la_0f_0,\\
\frac{\d f_N(t)}{\d t} &=&\la_{N-1}\,f_{N-1}-\mu_Nf_N,\nonumber
\eeqa
where $\la_k=u_{N-k}$ and $\mu_k=u_k$
are respectively the rate at which a particle enters site~1, coming
from site~2, or leaves site~1 for site~2.
The equations for $k=0$ or $k=N$ are special, since $u_0=0$.
The above equations describe a biased random walk on the interval $(0,N)$,
with reflecting boundaries at 0 and $N$,
the position of the walker being the random variable $N_1(t)$,
i.e., the number of particles on site~1.

We begin by a description of the statics.
The time-independent solution to~(\ref{2sites}) is easily seen to coincide with
the prediction of the general formula~(\ref{fkeq}) for the stationary
probabilities
\beq
f_{k,\eq}=\frac{p_k p_{N-k}}{Z_{2,N}},
\label{fkeq2s}
\eeq
where $p_k$ is given by~(\ref{pk}) and $Z_{2,N}$ by~(\ref{z0z1z2}).
Using the asymptotic form~(\ref{pkas}) of $p_k$
at large $k$, the probabilities~(\ref{fkeq2s}) are found to scale as
\be
f_{k,\eq}\approx\frac{C_N}{k^b(N-k)^b}
\ee
for large $k$ and $N-k$.
The amplitude $C_N$ is given below (see~(\ref{2sitb}) and~(\ref{2sitbb})).
The distribution profile is therefore U-shaped for any positive~$b$:
the most probable stationary configurations are those
where almost all the particles are located on one of the two sites.
In the language of the effective potential,
defined as $-\ln(f_{k,\eq}/f_{0,\eq})$,
these two states correspond to two symmetric valleys,
on both sides of a potential barrier situated at~$N/2$.
However, as we now discuss,
the cases $b>1$ and $0<b<1$ are qualitatively different.
They correspond respectively to the case of a high and a low
potential barrier (see Appendix B).

For $b>1$, the partition function $Z_{2,N}$
is dominated by configurations such that almost all the particles
lie on either of the sites, i.e., either $k$ is finite, or $N-k$ is finite.
Indeed the sum of the $p_k$,
i.e., $P(1)$ (see~(\ref{part})), is convergent.
We have
\beq
Z_{2,N}\approx2p_NP(1)\approx\frac{2b\Gamma(b+1)}{b-1}\,N^{-b},\qquad
C_N\approx\frac{(b-1)\Gamma(b)}{2}\,N^b.
\label{2sitb}
\eeq

For $0<b<1$, on the contrary, the sum of the $p_k$ is divergent.
As a consequence, all the values of $k$
contribute to the expression~(\ref{z0z1z2}) for the partition function
$Z_{2,N}$.
Evaluating this expression by means of an integral, we obtain
\beq\label{2sitbb}
Z_{2,N}\approx\frac{\Gamma(b+1)^2\Gamma(1-b)^2}{\Gamma(2-2b)}\,N^{1-2b},\qquad
C_N\approx\frac{\Gamma(2-2b)}{\Gamma(1-b)^2}\,N^{2b-1}.
\eeq

A natural definition of the flipping time is
the crossing time~$T=T_\L=T_\R$, i.e.,
the mean time taken by the walker to cross the system from right to left,
or from left to right.
As shown in Appendix B, the expression of $T$ only relies on
the knowledge of the stationary state.
It is given in terms of $f_{k,\eq}$ by~(\ref{sdef}),
from which analytical asymptotic estimates of $T$ are easily obtained.
Replacing the sum by an integral in~(\ref{sdef}), we have,
for $b>1$ (regime of high barrier),
\beq
T\approx\frac{b\Gamma(b+1)}{(b-1)\Gamma(2b+2)}\,N^{b+1},
\label{et1}
\eeq
while for $0<b<1$ (regime of low barrier),
\be
T\approx\frac{\pi b}{2(1-4b^2)\tan\pi b}\,N^2.
\ee

We complement this study by the determination of
the spectral properties of the master equation~(\ref{2sites}).
The latter reads formally
\beq
\frac{\d f_k(t)}{\d t}=\sum_\ell\M_{k,\ell}f_\ell(t),
\label{eq:markov}
\eeq
where $\M$ is the Markov matrix of the process.
Denoting its eigenvalues by $E_a$
and the corresponding eigenvectors by $\varphi_{k,a}$,
the eigenvalue problem for $\M$ consists in solving the equation
\beq
u_{k+1}\varphi_{k+1,a}+u_{N+1-k}\varphi_{k-1,a}-(u_k+u_{N-k})\varphi_{k,a}
=-E_a\,\varphi_{k,a},
\label{spectr2}
\eeq
obtained by looking for solutions to~(\ref{2sites}) of the form
$f_k(t)=\varphi_k\,\e^{-Et}$.
The ground state $E_0=0$ corresponds to the stationary solution
$\varphi_{k,0}=f_{k,\eq}$ of~(\ref{fkeq2s}).
The inverse of the other eigenvalues, $\tau_a=1/E_a$ for $a=1,\dots,N$,
are the characteristic times of the model.

Let us analyse the scaling behaviour of these times for $N$ large.
We make use of the results of Appendix~B
for the case where the potential has two symmetric valleys.

For $b>1$, the main characteristic time $\tau_1$ is related to the crossing
time~$T$ by~(\ref{tsym}):
\beq
\tau_1\approx\frac{T}{2}.
\label{tau1}
\eeq
It thus scales as $N^{b+1}$, with an explicitly known prefactor
(see~(\ref{et1})).
The subsequent characteristic times $\tau_2$, $\tau_3,\dots$,
diverge only as the diffusive timescale~$N^2$,
i.e., much less rapidly than $\tau_1$ for $b>1$.
This can be shown by taking the continuum limit of~(\ref{spectr2}).
Introducing the scaling variable $x=k/N$ and $\E=N^2E$,
we get the following differential equation
\beq
x^2(1-x)^2\varphi''+bx(1-x)(1-2x)\varphi'+[\E x^2(1-x)^2-b(1-2x+2x^2)]\varphi=0
\label{fcl}
\eeq
(a confluent Heun equation) for the scaling function $\varphi(x)$,
with Dirichlet boundary conditions: $\varphi(0)=\varphi(1)=0$.
Let $\E_a$ be the eigenvalues ($a=2,3,\dots$).
We have
\beq
\tau_a\approx\frac{N^2}{\E_a}.
\label{eta}
\eeq
A numerical integration of~(\ref{fcl}), with $b=2$, yields $\E_2=26.815$,
$\E_3=74.609$, and so on.

\begin{figure}[htb]
\begin{center}
\includegraphics[angle=90,width=.6\linewidth]{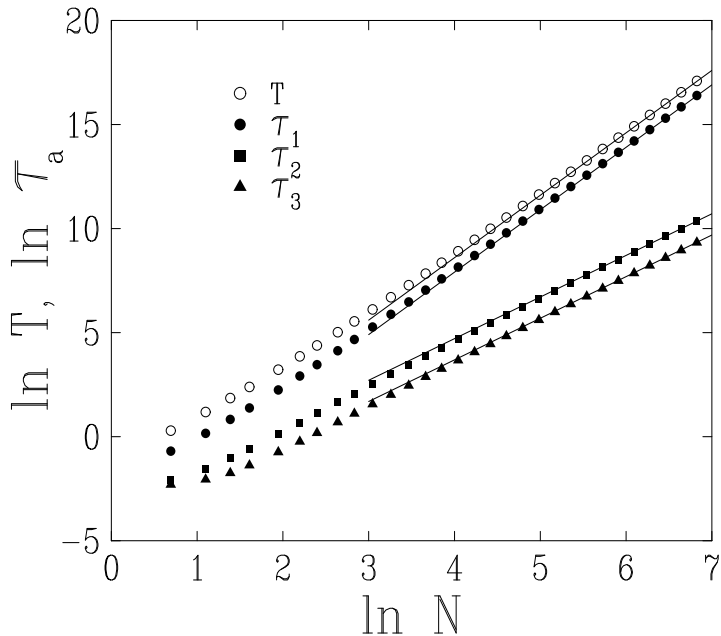}
\caption{\small
Log-log plot of the crossing time $T$
and of the first three characteristic times $\tau_1$, $\tau_2$, $\tau_3$,
against the number $N$ of particles, for the two-site problem with $b=2$.
Full straight lines: asymptotic scaling laws~(\ref{et1}),~(\ref{tau1}),
and~(\ref{eta}),
including prefactors: $T\approx N^3/30$, $\tau_1\approx N^3/60$,
$\tau_2\approx N^2/26.185$, $\tau_3\approx N^2/74.609$.}
\label{fe}
\end{center}
\end{figure}

Figure~\ref{fe} shows a log-log plot of the exact crossing time $T$,
as given by~(\ref{xtimes}),
and of the first three characteristic times $\tau_1$, $\tau_2$, $\tau_3$,
obtained from solving~(\ref{spectr2}) numerically,
against the number $N$ of particles, for $b=2$.
The agreement with the analytical asymptotic estimates~(\ref{et1}),
(\ref{tau1}), and~(\ref{eta}) is excellent.

For $0<b<1$, all the characteristic times, $\tau_1$, $\tau_2,\dots$,
grow proportionally to the diffusive timescale~$N^2$, as in~(\ref{eta}).

Consider finally the two-time correlation function for the occupation,
computed in the stationary state:
\be
C(t)=\mean{N_1(t)\,N_1(0)}-\mean{N_1}^2,
\ee
with $\mean{N_1}=N/2$.
This correlation function
is the simplest example of an observable falling off to zero
as $C(t)\sim\exp(-t/\tau_1)$,
where $\tau_1=1/E_1$ is the main characteristic time.
Indeed, in the present case,
the temporal evolution of $C(t)$ follows the Markovian scheme presented above.
We have
\be
\mean{N_1(t)\,N_1(0)}=\sum_{j,k\ge1}
j\,k\,\P(N_1(t)=k|N_1(0)=j)\,\P(N_1(0)=j),
\ee
i.e.,
\be
\mean{N_1(t)\,N_1(0)}=\sum_{j,k\ge1}j\,k\,p_{j,k}(t)\,f_{j,\eq}
=\sum_{k\ge1}k\,\gamma_k(t),
\ee
where the quantities
\be
p_{j,k}(t)=\P(N_1(t)=k|N_1(0)=j),\qquad
\gamma_k(t)=\sum_{j\ge1}j\,\,p_{j,k}(t)\,f_{j,\eq}
\ee
follow the master equation~(\ref{2sites}), with initial conditions
\be
p_{j,k}(0)=\delta_{j,k},\qquad\gamma_k(0)=k\,f_{k,\eq},
\ee
consistently with $C(0)=\sum_k k^2\,f_{k,\eq}-\mean{N_1}^2$.

To summarise, the case of two sites is instructive in several respects.
The model exhibits two qualitatively different dynamical regimes.
For $b>1$, the dynamics is an example of a diffusion in a potential with a high
barrier.
The flipping time, namely the crossing time $T$,
diverges as $N^{b+1}$, as does the main characteristic time $\tau_1$.
The limit value of the ratio $T/\tau_1$ is equal to 2, characteristic
of a symmetric potential with a high barrier.
For $0<b<1$, the flipping time, as well as all characteristic times,
now grow as the diffusive timescale $N^2$.
The ratio $T/\tau_1$ is non trivial and depends on $b$.
Appendix~B treats the general case of a non-symmetric potential,
for use in the next section.

\section{The general case: Markovian Ansatz}
\label{secma}

We now proceed to the analysis of the stationary dynamics of the condensate
for the general case ($M$ and $N$ arbitrary, $b>2$, $\rho>\rho_c$).
We first make a crucial observation on the behaviour of the occupation
probabilities $f_{k,\eq}$
in the region $k\gg1$ and $\D-k\gg1$, hereafter referred to as the region of
the probability ``dip''.
We refer the reader to~\cite{maj,evans2} for a general study of the statics of
mass transport models
with critical background, for a finite system,
as well as to the additional information contained in Appendix~A.

\begin{figure}[htb]
\begin{center}
\includegraphics[angle=90,width=.6\linewidth]{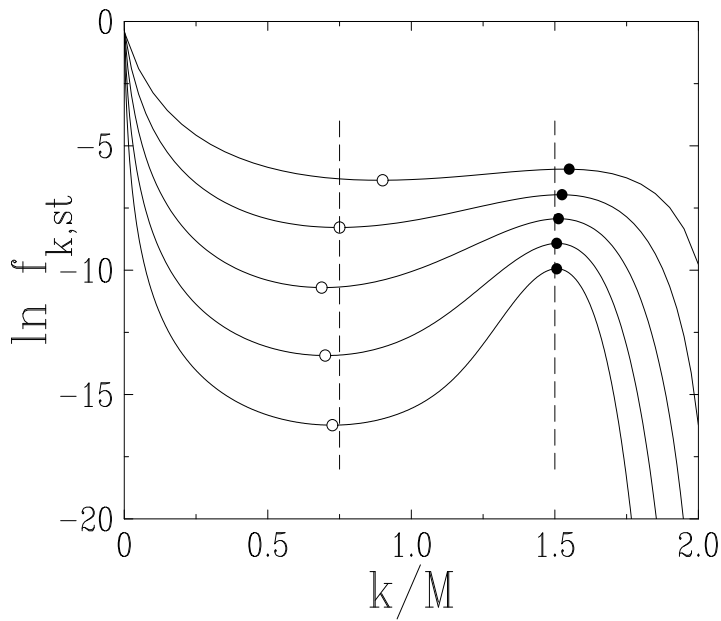}
\caption{\small
Logarithmic plot of the occupation probabilities $f_{k,\eq}$
in the condensed phase ($b=4$, $\rho=4\rho_c=2$), against the ratio $k/M$.
Top to bottom: $M=20$, 40, 80, 160, and 320.
Full lines: $f_{k,\eq}$ obtained by~(\ref{zrecrel}) and~(\ref{afkeq}).
Full (empty) symbols: maxima (minima) of occupation probability.
Dashed vertical lines:
asymptotic locations of the minima: $k/M=\D/(2M)=(\rho-\rho_c)/2=3/4$,
and of the maxima: $k/M=\D/M=\rho-\rho_c=3/2$.}
\label{ffk}
\end{center}
\end{figure}

Figure~\ref{ffk} shows a logarithmic plot of the occupation probabilities
$f_{k,\eq}$ in the condensed phase,
computed using equations~(\ref{zrecrel}) and~(\ref{afkeq}),
against the ratio $k/M$, for $b=4$, $\rho=2$, and several values of $M$.
This plot exhibits the following features.
For $k\ll M$, the distribution~$f_{k,\eq}$ is approximately given
by the power law~(\ref{fkc}) of an infinite critical system.
The contribution of the condensate slowly builds up as a probability hump
around $k=\D$, the mean number of excess particles.
Finally one observes a broad and shallow probability dip in the region located
between the critical background and the condensate hump.
It is shown in Appendix~A that the region of the dip is dominated by
configurations
where the excess particles are shared by {\it two} sites (see~(\ref{afkdip})):
\beq
f_{k,\eq}\approx(b-1)\Gamma(b)\frac{\D^b}{k^b(\D-k)^b}.
\label{fkdip}
\eeq
The observed locations of the maxima ($k\approx\Delta$) and minima
($k\approx\Delta/2$) of the occupation probabilities
corroborate this picture, as explained in the caption of Figure~\ref{ffk}.

These observations lead us to a first crude estimate
of the characteristic time of the stationary-state dynamics of the condensate.
The corresponding decay rate can indeed be argued to be proportional
to the minimum $f_\min$ of the stationary probabilities $f_{k,\eq}$.
We thus obtain
\beq
\tau_\a\sim\frac{1}{f_\min}.
\label{arrmin}
\eeq
We shall refer to this kind of estimate as the Arrhenius law,
as it generalises the formula originally due to Arrhenius~\cite{ar}
for a chemical reaction rate in terms of the corresponding activation energy.
In the present context, the result~(\ref{fkdip})
implies that $f_\min$ is reached near the middle of the dip region
$(k\approx N/2)$, and that the Arrhenius law~(\ref{arrmin}) yields
\beq
\tau_\a\sim\D^b.
\label{arrdip}
\eeq

We now turn to the derivation of a more refined approximation,
yielding the correct prefactor to the Arrhenius law,
somewhat along the lines of the approach initiated by Kramers~\cite{kr}
for chemical reactions.
The above observations on the statics suggest to propose an
effective description of the stationary
dynamics of the condensate, based on a {\it Markovian Ansatz}.
Assume that the condensate is on site number 1
at the initial observation time ($t=0$).
The number $N_1(t)$ of particles on that site is initially
very large, $N_1(0)\approx\D$, and therefore evolves slowly,
until the condensate dissolves into the critical background.
Thus

\begin{itemize}
\item
We single out $N_1(t)$ as the collective co-ordinate of the system,
that is the appropriate slow variable describing the dynamics of the condensate.
\item
We model the dynamics of $N_1(t)$ by~(\ref{rat}),
i.e., by a biased diffusive motion on the interval $k=0,\dots,N$.
The left hopping rates are taken equal to the microscopic ones: $\mu_k=u_k$.
The right hopping rates $\la_k$ are so chosen that, in the stationary state,
the probabilities $f_{k,\eq}$ of the effective model coincide with the
occupation probabilities~(\ref{fkeq}) of the original ZRP.
The detailed balance condition~(\ref{deba}) yields
\be
\la_k=\frac{\mu_{k+1}f_{k+1,\eq}}{f_{k,\eq}}=\frac{Z_{M-1,N-k-1}}{Z_{M-1,N-k}}.
\ee
The rate $\la_k$ is thus a function of $k$, $M$, and $N$.
For $M=2$, this formula gives $\la_k=u_{N-k}$, as expected.
In the fluid phase, in the thermodynamic limit,
the rates~$\la_k$ converge to $z_0$, defined in~(\ref{col}).
Finally, in the condensed phase, in the dip region, we obtain
\be
\la_k\approx 1+\frac{b}{\D-k}.
\ee
\end{itemize}

This effective description therefore reduces the full model to a
Markovian model for one degree of freedom in an asymmetric potential.
The two valleys of the potential are separated by a high (power-law) barrier.
The left potential valley, corresponding to the critical background,
has a weight $P_\L\approx1$, whereas the right potential valley,
corresponding to the hump of the condensate, has a weight $P_\R\approx1/M\ll1$.

In this framework (see Appendix B) the stationary dynamics
of the condensate is characterised by a single diverging timescale.
We choose to {\it define} this timescale, denoted by~$\tau_\ma$,
to be the crossing time~$T_\L$ from the right valley to the left one
in the effective Markovian problem.
The characteristic time is thus expressed by~(\ref{xtimes}), i.e.,
\beq
\tau_\ma=T_\L=\sum_{\ell=1}^N\frac{1}{\mu_\ell f_{\ell,\eq}}
\sum_{m=\ell}^Nf_{m,\eq},
\label{tauma}
\eeq
in terms of known quantities, the rates $\mu_k$
and the stationary probabilities $f_{k,\eq}$.

Its asymptotic growth is easily determined by noting that~(\ref{tauma})
is dominated by the behaviour of the probability $f_{k,\eq}$ in the region of
the dip.
Hence, inserting the expression~(\ref{fkdip}) into~(\ref{tauma}),
and evaluating the sum as an integral, we obtain
\beq
\tau_\ma\approx\frac{b\Gamma(b+1)}{(b-1)\Gamma(2b+2)}\,\frac{\D^{b+1}}{M}
=\frac{b\Gamma(b+1)}{(b-1)\Gamma(2b+2)}\,(\rho-\rho_c)^{b+1}M^b.
\label{zrpt1}
\eeq
This result incorporates the full prefactor
to the Arrhenius estimate~(\ref{arrdip}),
including a factor of $\D/M=\rho-\rho_c$,
as well as an absolute numerical factor.
It is also worth noticing the exact correspondence between~(\ref{zrpt1})
and the formula~(\ref{et1}) for the two-site system,
up to the replacement of $M$ by 2 and of $\D$ by $N$,
and up to an absolute factor 2.

\begin{figure}[htb]
\begin{center}
\includegraphics[angle=90,width=.6\linewidth]{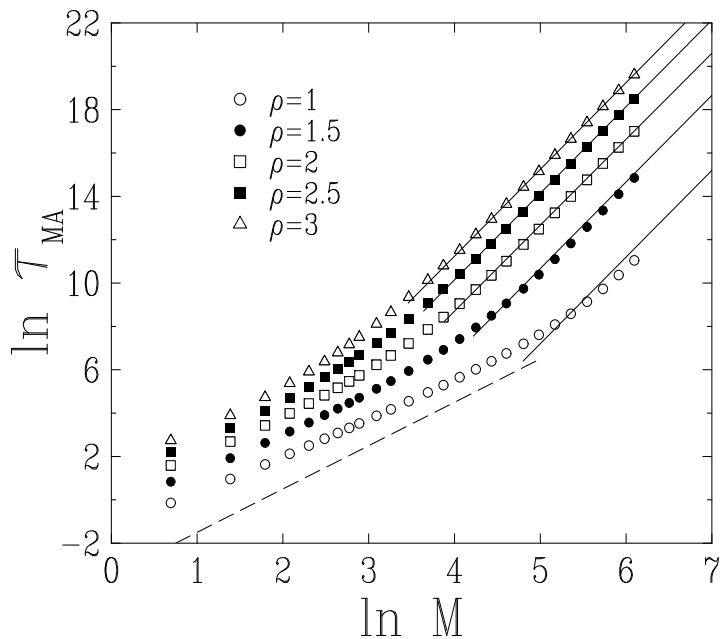}
\caption{\small
Log-log plot of the prediction $\tau_\ma$ of the Markovian Ansatz
for the characteristic time of condensate dynamics,
against the system size $M$, for $b=4$ (hence $\rho_c=1/2$)
and several values of the density in the condensed phase.
Symbols: exact values of $\tau_\ma$, obtained by means of~(\ref{tauma}).
Full straight lines:
asymptotic scaling result~(\ref{zrpt1}), including prefactor.
Dashed straight line: diffusive timescale of the critical dynamics.}
\label{fanal}
\end{center}
\end{figure}

Figure~\ref{fanal} shows a log-log plot of the characteristic time
$\tau_\ma$, given by~(\ref{tauma}), against the system size $M$,
for $b=4$ and several values of the density in the condensed phase.
The scaling form~(\ref{zrpt1}) only sets in for rather large systems.
For $\rho-\rho_c$ small, there is indeed a crossover phenomenon
between the diffusive scaling of the timescale at the critical point,
$\tau_\ma\sim M^2$, irrespective of $b$,
and that of the condensed phase, $\tau_\ma\sim(\rho-\rho_c)^{b+1}M^b$.
The crossover takes place for a system size
of order $M_\star\sim(\rho-\rho_c)^{-(b+1)/(b-2)}$.
The apparent crossover size $M_\star$,
beyond which the form~(\ref{zrpt1}) holds,
remains however rather large,
even when the density $\rho$ largely exceeds~$\rho_c$.
For instance, for $b=4$ and $\rho=4\rho_c=2$
(the values used in the Monte-Carlo
simulations of the next section), one has $M_\star\approx40$.

It is also worth considering $T_\R$, the other crossing time
of the effective Markovian problem.
First of all, the estimate~(\ref{tltr}) for $T_\R$ fails
in the present situation.
Indeed, this estimate is derived in the framework considered in Appendix~B,
where the effective potential is assumed to be small everywhere except
in the barrier.
The present form of the right valley is however very different.
The stationary probabilities $f_{k,\eq}$
indeed fall off very fast as $k$ goes to its maximal value~$N$,
so that the effective potential $V_k$ exhibits a sudden rise as $k\to N$.
As a consequence, for a large system size,
the exact formula~(\ref{xtimes}) for~$T_\R$ is not dominated
by the region of the dip.
It is instead entirely dominated
by the last stationary probability, which reads $f_{N,\eq}=p_N/Z_{M,N}$.
Using the expression~(\ref{zas}) for $Z_{M,N}$, we get
\be
T_\R\approx\frac{1}{f_{N,\eq}}\approx M\left(\frac{\rho}{\rho-\rho_c}\right)^b
\left(\frac{b}{b-1}\right)^{M-1}.
\ee
The crossing time $T_\R$ therefore grows exponentially with the system size $M$.
The corresponding activation energy reads
$-\ln f_{N,\eq}=V_N\approx M\ln(b/(b-1))=-M\F_c$,
where $\F_c$ is the critical free energy density.
This crossing time has a clear physical interpretation:
it is the characteristic time of the occurrence
of the very improbable event where the whole critical background of the ZRP
collapses onto the condensate (whence the occurrence of $\F_c$).
The occupation of the condensate is then equal to $N$,
well above the usual $\Delta$.
The exponentially large timescale $T_\R$
is absent from the spectrum of the Markov matrix of the effective description,
whose first eigenvalue scales as $E_1\approx1/(P_L T_L)\approx1/T_L$.
The occurrence of exponentially large first-passage times
has been underlined in several instances of urn models~\cite{barc,activ}.

\section{The general case: numerical simulations}
\label{secsim}

In order to test the predictions of the Markovian Ansatz,
we now compare them to results of numerical simulations.
We have performed Monte-Carlo simulations of the ZRP with the
rates~(\ref{rates}), using random sequential updates,
in the three geometries defined above:
mean-field (MF), one-dimensional symmetric (1DS) ($p=q=1/2$),
and one-dimensional asymmetric (1DAS) ($p=1$).
We have focussed our attention onto the following stationary-state correlations,
which are devised to be especially sensitive to the
dynamics of the condensate:

\begin{itemize}
\item
Local density correlation function:
\be
C_\loc(t)=\mean{N_1(t)N_1(0)}-\mean{N_1}^2,
\ee
where $\mean{N_1}=\rho$.
The initial value $C_\loc(0)=\sum_{k=0}^Nk^2\,f_{k,\eq}-\rho^2$
reads approximately $C_\loc(0)\approx M(\rho-\rho_c)^2=\D^2/M$.
\item
First harmonic of global density fluctuations:
\be
C_\gl(t)=\frac{1}{M}\,\mean{g(t)g^\star(0)},
\ee
where
\be
g(t)=\sum_{m=1}^N N_m(t)\,\e^{2\pi\i m/M}
\ee
is the first Fourier coefficient of the density fluctuations.
This definition only makes sense for the one-dimensional geometry
with periodic boundary conditions.
\end{itemize}

Both correlation functions fall off to zero in the limit of long times,
by construction.
The correlation functions measured in simulations are found to
obey a very clear exponential fall-off, except for very small systems,
confirming thus the emergence of a single well-defined slow timescale
characterising the dynamics of the condensate.
In the one-dimensional geometries (1DS and 1DAS),
the local and global density correlations have different relaxation times:
\be
C_\loc(t)\sim\exp(-t/\tau_\loc),\qquad C_\gl(t)\sim\exp(-t/\tau_\gl).
\ee
In the mean-field geometry only $\tau_\loc$ can be measured.

\begin{figure}[htb]
\begin{center}
\includegraphics[angle=90,width=.6\linewidth]{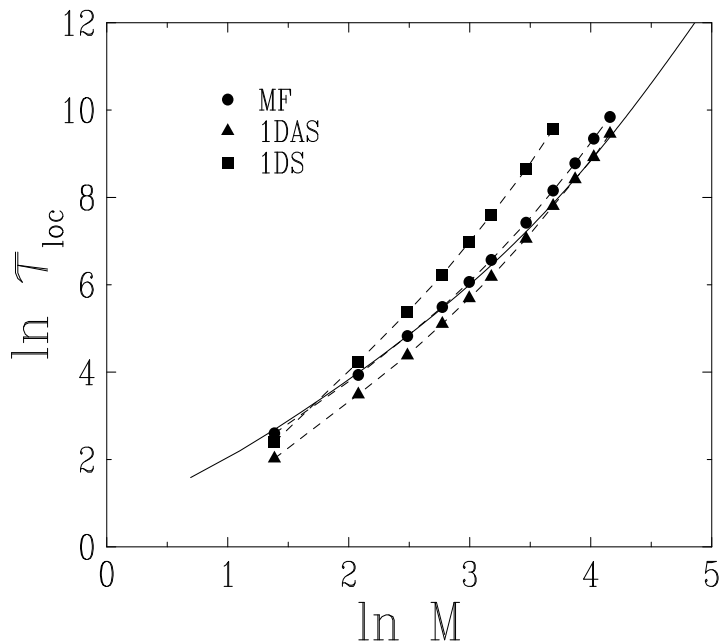}
\caption{\small
Log-log plot of the relaxation time $\tau_\loc$
measured from the fall-off of the local density correlation $C_\loc(t)$,
against the system size $M$ ($b=4$, $\rho=4\rho_c=2$).
Symbols: numerical data for the three geometries considered.
Full line: prediction $\tau_\ma$ of the Markovian Ansatz.}
\label{fsim}
\end{center}
\end{figure}

Figure~\ref{fsim} demonstrates that the prediction of the Markovian Ansatz
correctly describes the main dependence
of the measured relaxation time $\tau_\loc$ of local density fluctuations
(the relaxation time $\tau_\gl$ has been omitted for clarity).
The agreement is especially visible in the MF and 1DAS models,
whereas the relaxation time of the 1DS model grows slightly faster.
The prediction $\tau_\ma$ of the Markovian Ansatz
therefore provides a good estimate of the characteristic time
of the dynamics of the condensate, even for moderately large system sizes,
below the crossover size $M_\star$
where the power-law scaling result~(\ref{zrpt1}) sets in.

In order to analyse in more detail the behaviour of the relaxation times
$\tau_\loc$ and $\tau_\gl$ of local and global density fluctuations,
we introduce the dimensionless time ratios
\beq
R_\loc=\frac{\tau_\loc}{\tau_\ma},\qquad R_\gl=\frac{\tau_\gl}{\tau_\ma}.
\label{timer}
\eeq

\begin{figure}[htb]
\begin{center}
\includegraphics[angle=90,width=.6\linewidth]{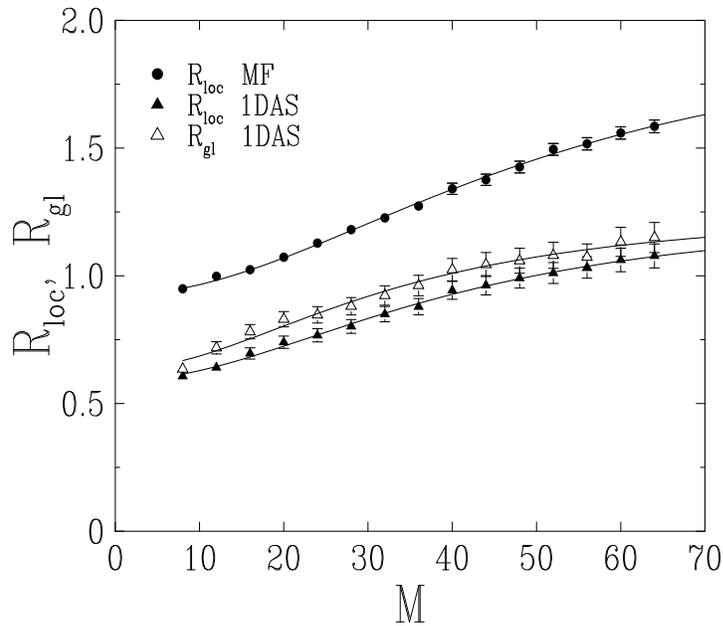}
\caption{\small
Plot of the relaxation time ratios $R_\loc$ and $R_\gl$ introduced
in~(\ref{timer}), against the system size $M$, for the MF and 1DAS geometries
($b=4$, $\rho=4\rho_c=2$).
Symbols with error bars: numerical data.
Full lines: Rational fits (see text)
leading to the limiting values $R_\loc^\infy\approx2$ (MF)
and $R_\loc^\infy\approx R_\gl^\infy\approx1.26$ (1DAS).}
\label{fr}
\end{center}
\end{figure}

Figure~\ref{fr} shows a plot of the time ratio $R_\loc$ in the MF model,
and of both time ratios $R_\loc$ and $R_\gl$ in the 1DAS model.
The three series of data remain of order unity,
but exhibit a weak residual growth as a function of $M$,
at least in the range of system sizes shown on the Figure, which
slows down beyond system sizes
of the order of the crossover size $M_\star\approx40$,
and seems to eventually saturate.
Let us stress the difficulty in measuring these time ratios with
a statistical error of a few percent for larger systems,
whereas the measured relaxation times vary over four orders of magnitude.
The fact that the time ratios saturate to finite values
in the thermodynamic limit is further supported
by the fits of the data by fractional linear functions
of $1/M^2$, shown as full lines on Figure~\ref{fr}.
The rationale behind this choice of a fitting function is the following.
For $b=4$, $\tau_\ma$ scales as $M^4$, whereas subleading terms
proportional to the diffusive timescale $M^2$ are expected,
among other correction terms.
The presented fits lead to the following limiting values of the time ratios:
$R_\loc^\infy\approx2$ (MF) and $R_\loc^\infy\approx R_\gl^\infy\approx1.26$
(1DAS), hence
\beq
\tau_\loc\approx2\,\tau_\ma\quad\hbox{(MF)},\qquad
\tau_\loc\approx\tau_\gl\approx1.26\,\tau_\ma\quad\hbox{(1DAS)}.
\label{tinfy}
\eeq
The scaling formula~(\ref{zrpt1}) of the Markovian Ansatz prediction
therefore correctly describes the asymptotic behaviour of the relaxation times
in the MF and 1DAS cases, up to finite prefactors.

\begin{figure}[htb]
\begin{center}
\includegraphics[angle=90,width=.6\linewidth]{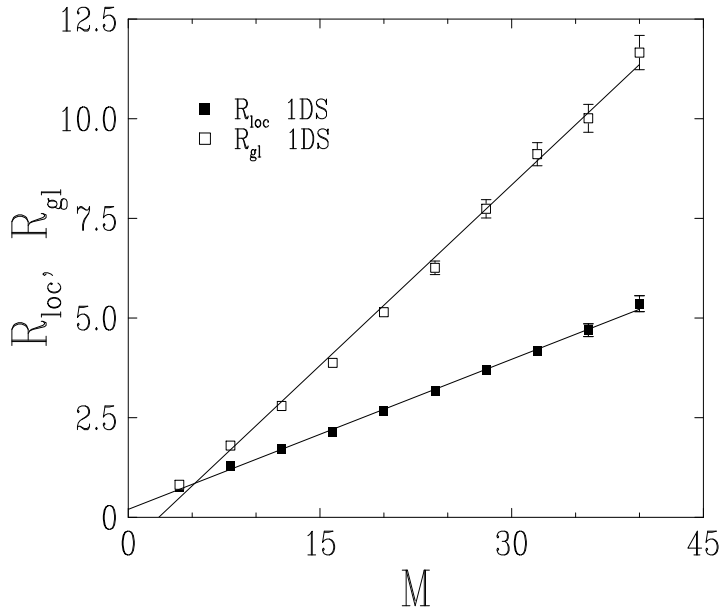}
\caption{\small
Plot of the relaxation time ratios $R_\loc$ and $R_\gl$ introduced
in~(\ref{timer}), against the system size $M$, for the 1DS geometry
($b=4$, $\rho=4\rho_c=2$).
Symbols with error bars: numerical data.
Full straight lines: least-square fits with slopes 0.12 and 0.30.}
\label{frs}
\end{center}
\end{figure}

Figure~\ref{frs} shows plots of the time ratios $R_\loc$ and $R_\gl$
in the 1DS geometry.
Both ratios exhibit a clear linear dependence on the system size $M$,
implying that the relaxation times of local and global density fluctuations
scale as
\beq
\tau_\loc\approx0.12\,M\tau_\ma,
\qquad\tau_\gl\approx0.30\,M\tau_\ma\quad\hbox{(1DAS)}.
\label{tsinfy}
\eeq
These expressions of the relaxation times in the 1DS geometry
therefore contain an extra factor of the system size $M$,
with respect to the analytical prediction of the Markovian Ansatz,
and to the results in the MF and 1DAS geometries.
This factor originates in the Gambler's ruin problem~\cite{ruin}
(see Discussion).
To close up this part on numerical results,
let us emphasise that the prefactors entering~(\ref{tinfy}) and~(\ref{tsinfy})
have been extracted accurately only for $b=4$ and $\rho=2$.
These prefactors are however expected to have a weak dependence
on $b$ and $\rho$.

\section{Discussion}

This paper is devoted to the dynamics in the stationarity state
of the macroscopic condensate
occurring in a class of nonequilibrium statistical mechanical models.
The problem is well-defined for a finite system
consisting of $M$ sites, where the condensate is made of
a large but finite number $\D=M(\rho-\rho_c)$
of excess particles, up to fluctuations.

In this work, we have introduced, for the first time, a description
of the stationary-state dynamics of the condensate, on the example of a class
of ZRP.
This effective description, based on a Markovian Ansatz,
relies deeply on the exact knowledge of the stationary-state
occupation probabilities $f_{k,\eq}$ of the finite system.
The essential feature of these probabilities
is the occurrence of a dip region dominated by configurations
where the excess particles are shared by two sites.
The effective Markovian problem thus derived is an example of biased diffusion
in the presence of a bistable potential with a high barrier separating two
valleys.
The present approach turns the Arrhenius estimate $\tau_\a\sim1/f_\min$
of the characteristic time into the accurate prediction $\tau_\ma$,
just as the method initiated by Kramers improves the Arrhenius law
in the context of chemical reactions.
Our main result is the following.
{\it The characteristic timescale for the dynamics
of the condensate,~$\tau_\ma$, always grows much faster
than the diffusive time\-scale~$M^2$, but not exponentially~fast.}
For the class of ZRP with static exponent~$b>2$,
the characteristic time $\tau_\ma$ grows as~$M^b$.

These theoretical predictions are corroborated by the results
of Monte-Carlo simulations in the mean-field (MF), one-dimensional asymmetric
(1DAS), and one-dimensional symmetric (1DS) geometries.
Local and global temporal correlations of density fluctuations
in the stationary state have been respectively measured
by the correlation functions $C_\loc(t)$ and $C_\gl(t)$.
The corresponding relaxation times $\tau_\loc$ and $\tau_\gl$
are found to scale as follows:
$\tau_\loc\sim\tau_\ma\sim M^b$ (MF),
$\tau_\loc\sim\tau_\gl\sim\tau_\ma\sim M^b$ (1DAS),
and $\tau_\loc\sim\tau_\gl\sim M\tau_\ma\sim M^{b+1}$ (1DS),
up to absolute prefactors of order unity,
which are expected to bear a weak dependence on $b$ and $\rho$.
The local and global relaxation times seem to be asymptotically
equal ($\tau_\gl/\tau_\loc\approx1$) in the 1DAS geometry,
and asymptotically proportional ($\tau_\gl/\tau_\loc\approx2.5$)
in the 1DS geometry.
This observation confirms the visual impression gained from Figure~\ref{fhist}:
the stationary dynamics is non-local, in the sense that the jumps
of the condensate spread more or less uniformly over the whole sample.
(Any more local kind of dynamics,
where large jumps are either absent or improbable,
would imply that the ratio $\tau_\gl/\tau_\loc$ diverges with the system size.)

It is of interest to put our results on the dynamics of the condensate in the
stationary state in perspective with those of the non-stationary dynamics,
that we first recall.
The non-stationary dynamics of the formation of the condensate
has been investigated in some detail for ZRP, both in the mean-field
geometry~\cite{cg} and in one dimension~\cite{cg,gross}.
The following scenario holds in the MF and 1DAS geometries.
In the first stage of the dynamics, a coarsening phenomenon takes place:
the typical occupation of the most populated sites grows as $t^{1/2}$,
whereas the number of those sites decays as $M/t^{1/2}$.
After a time of order $M^2$, the system contains a finite number
of highly populated sites, i.e., condensate precursors.
The second stage of the non-stationary dynamics,
where all but one of the precursors die out,
also lasts a length of time of the order of the diffusive timescale $M^2$.
The whole non-stationary process of the formation of the condensate
is therefore characterised by a single timescale $\tau_\noneq\sim M^2$.
A similar scenario holds in the 1DS geometry,
the only difference being that $\tau_\noneq$ now scales as $M^3$.
Table~\ref{one} summarises the values of the dynamical exponents $z$ and $Z$,
such that $\tau_\noneq\sim M^z$ and $\tau\sim M^Z$,
where $\tau\sim\tau_\loc\sim\tau_\gl$
is the characteristic timescale for the stationary motion of the condensate.
The shift of the dynamical exponent by one unit in the 1DS geometry
has a common origin~\cite{cg,gross}: it stems from the Gambler's ruin
problem~\cite{ruin}.
An analogous phenomenon is encountered for example in the coarsening law
for the domain growth, and
in the motion of a tagged particle, in 1D Kawasaki dynamics~\cite{kawa}.

\begin{table}[ht]
\begin{center}
\begin{tabular}{|c|c|c|}
\hline
Geometry&$z$&$Z$\\
\hline
MF, 1DAS&2&$b$\\
\hline
1DS&3&$b+1$\\
\hline
\end{tabular}
\caption{\small Non-stationary and stationary dynamical exponents
of the ZRP with static exponent $b>2$.}
\label{one}
\end{center}
\end{table}

As recalled above, the non-stationary dynamical exponents
are insensitive to the exponent $b$, and more generally to the statics,
provided the system is in its condensed phase.
This feature is easily understood in the context of the
Markovian Ansatz proposed in the present work.
The last stage of the formation of the condensate,
i.e., the disap\-pearance of the smaller of the last two precursors,
implies no barrier crossing.
In terms of the occupation of the condensate,
it indeed corresponds to the transition from $N_1$ to $\D$,
where the initial occupation $N_1$ of the larger precursor
was already larger than $\D/2$,
corresponding to the top of the potential barrier.
This explains why $\tau_\noneq$ is given by the diffusive timescale,
both in the framework of the Markovian Ansatz
and in the MF and 1DAS geometries.

The prefactors appearing in~(\ref{tinfy}) demonstrate that
the dynamics is faster in the 1DAS geometry than in the MF geometry
by a factor of order $2/1.26\approx1.59$.
This effect can be explained at least qualitatively in terms of the mean time
$T_\mimi$ taken by a particle to travel between two randomly chosen sites.
In the 1DAS geometry, the motion is ballistic, implying that time is just
distance, hence $T_\mimi=M/2$.
In the MF geometry it can be shown that $T_\mimi=M$.
The factor 2 between these two estimates provides a rough
explanation of the observed ratio of characteristic times.

The present approach can be easily generalised to other classes of ZRP.
One situation of interest is when the critical occupation probabilities
fall off more rapidly than any power law (so that formally $b=\infty$).
Using both the expression~(\ref{gfkdip}) of the probabilities in the dip region,
and the Arrhenius law~(\ref{arrmin}) (or~(\ref{arr})) for barrier crossing,
we predict that the characteristic time scales as
\beq
\tau_\ma\sim\frac{1}{f_{\D/2}}\sim\frac{p_\D}{{p_{\D/2}}^2}.
\label{tauexp}
\eeq
This estimate diverges faster than any power of $M$,
but not exponentially fast.
The power-law prefactor, which is missed
by using the Arrhenius expression $\tau_\a$ instead of the full
prediction $\tau_\ma$, is therefore irrelevant.
An explicit example is provided by the rate~(\ref{grates}).
The critical occupation probabilities decay as
the stretched exponential~(\ref{gdecay}), whereas the characteristic time
has a stretched exponential dependence in the system size $M$:
\be
\tau_\ma\sim\exp\left(\frac{a}{1-\s}\left(2^\s-1\right)(\rho-\rho_c)^{1-\s}
M^{1-\s}\right).
\ee

The predictions obtained in this work
can also be extended to other geometries, besides the MF and 1D cases,
in analogy with the case of the non-stationary dynamics
of formation of the condensate~\cite{dgg}.
The MF exponents (first line of Table~\ref{one})
are expected to hold for partially or totally asymmetric dynamics
in any dimension, and for symmetric dynamics in dimension $d>2$.
Finally, a logarithmic factor is expected for symmetric dynamics
in two dimensions: $\tau\sim\tau_\ma\ln M\sim M^b\ln M$.

As a last remark, let us mention that it would be interesting to see whether
the predictions made in the present work extend to
the case of driven diffusive systems leading to condensation.
If, as currently believed, these systems can be adequately mapped onto ZRP in
the stationary state~\cite{wis1}, one would expect this to hold.

We end up with a critical review of references~\cite{gross,maj},
where the stationary-state motion of the condensate had already been addressed.
Reference~\cite{gross} contains predictions both on
the late stage of the formation of the condensate (non-stationary dynamics),
and on the dynamics of the condensate in the stationary state.
These predictions rely upon the analysis of an
effective symmetric two-site model.
In the stationary state this model leads to
an inverse binomial distribution.
(By means of an argument equivalent to the Arrhenius law,
the characteristic time for a macroscopic fluctuation
of the condensate is then found to grow exponentially with the system size.)
In our language, reference~\cite{gross} predicts that the stationary
occupation probabilities are exponentially small in the dip region,
which is in strong disagreement
with the power-law scaling result~(\ref{fkdip}),
itself a direct consequence of the exactly known finite-size
stationary probabilities~(\ref{fkeq}).
The effective model proposed in~\cite{gross} is therefore inappropriate
to describe the stationary ergodic motion of the condensate.
(One possible explanation is that the rates defining the effective model
are derived under an implicit assumption of weak fluctuations
in the probability currents, which might not hold when the condensate jumps
from one site to another.)
Reference~\cite{maj} is devoted to the statics of finite-size systems
for a class of one-dimensional mass transport models leading to condensation.
The authors also give an estimate for the characteristic time $\tau$ of the
dynamics of the condensate at stationarity.
This time is argued to diverge as some power of the system size
in the regime of anomalous fluctuations ($2<b<3$)
and exponentially fast in the regime of normal fluctuations ($b>3$).
This prediction is incorrect because the required probability is estimated
by using an expression for the condensate hump outside its range of validity.

\subsection*{Acknowledgements}

It is a pleasure to thank J.~Houdayer for stimulating discussions
on numerical simulations,
and M.~Evans, S.~Grosskinsky, S.~Majumdar, and G.~Sch\"utz
for exchange of correspondence.

\appendix
\setcounter{equation}{0}
\def\theequation{A.\arabic{equation}}
\section*{Appendix~A.~Statics of the ZRP}

This Appendix presents a reminder of the main static properties of the ZRP,
and a derivation of a few results relevant to section 5.

\subsection*{Simple facts}

The partition function $Z_{M,N}$ entering~(\ref{eq:pn}) reads
\be
Z_{M,N}=\sum_{\{N_i\}}\prod_{i=1}^M p_{N_i}\,
\delta\left(\sum_{i=1}^M N_i,N\right),
\ee
and obeys the recursion formula
\beq
Z_{M,N}=\sum_{k=0}^N p_k\,Z_{M-1,N-k}.
\label{zrecrel}
\eeq
This ensures that the stationary occupation probabilities
\beq
f_{k,\eq}=\frac{p_k\,Z_{M-1,N-k}}{Z_{M,N}}
\label{afkeq}
\eeq
are normalised.
We have
\beq
Z_{0,N}=\delta_{N,0},\qquad
Z_{1,N}=p_N,\qquad
Z_{2,N}=\sum_{k=0}^N p_kp_{N-k},
\label{z0z1z2}
\eeq
and so on.
Using an integral representation of the Kronecker delta function, we obtain
\beq
Z_{M,N}=\oint\frac{\d z}{2\pi\i z^{N+1}}\,P(z)^M,
\label{contour}
\eeq
where the generating series of the weights $p_k$ reads
\be
P(z)=\sum_{k\ge0} p_kz^k.
\ee
Static properties of the ZRP are therefore entirely encoded in this series.
In the thermodynamic limit ($M\to\infty$ at fixed density $N/M=\rho$),
the free energy per site,
\be
\F=-\lim_{M\to\infty}\frac{1}{M}\ln Z_{M,N},
\ee
can be obtained by evaluating the
contour integral in~(\ref{contour}) by the saddle-point method.
The saddle-point value $z_0$ depends on the density $\rho$ through the equation
\beq
\frac{z_0 P'(z_0)}{P(z_0)}=\rho.
\label{col}
\eeq
The free energy per site and the stationary occupation probabilities read
\be
\F=\rho\ln z_0-\ln P(z_0),\qquad f_{k,\eq}=\frac{p_k\,z_0^k}{P(z_0)}.
\ee

\subsection*{Rates $u_k=1+b/k$: Power-law critical behaviour}

For the choice of rate $u_k=1+b/k$, the above formulas yield
\beqa\label{pkas}
&&p_k=\frac{\Gamma(b+1)\,k!}{\Gamma(k+b+1)}
=\int_0^1 u^k\,b(1-u)^{b-1}\,\d u\approx\frac{\Gamma(b+1)}{k^b},\nonumber\\
&&P(z)=\int_0^1\frac{b(1-u)^{b-1}}{1-zu}\,\d u=\null_2F_1(1,1;b+1;z),
\eeqa
where $_2F_1$ is the hypergeometric function.
The function $P(z)$ has a branch cut at $z=1$, with a singular part of the form
\be
P_\sg(z)\approx A\,P(1)(1-z)^{b-1},
\ee
so that $P(z)$ is only differentiable $\Int(b)-1$ many times at $z=1$.
The following values are of interest:
\beq
\matrix{
P(1)=\frad{b}{b-1},\hfill&A=\frad{(b-1)\pi}{\sin\pi b},\hfill\cr
P'(1)=\frad{b}{(b-1)(b-2)},\quad&P''(1)=\frad{4b}{(b-1)(b-2)(b-3)}.}
\label{part}
\eeq
Whenever $b=n\ge2$ is an integer, the amplitude $A$ diverges.
The singular part of the generating series is of the form
$P_\sg(z)\approx n(-1)^n(1-z)^{n-1}\ln(1-z)$.

For $b>2$, the system has a continuous phase transition
at a finite critical density
\be
\rho_c=\frac{P'(1)}{P(1)}=\frac{1}{b-2},
\ee
such that the saddle point $z_0$ reaches the singular point $z=1$.
This critical density separates a fluid phase $(\rho<\rho_c)$
and a condensed phase $(\rho>\rho_c)$.

\subsubsection*{Critical density $(\rho=\rho_c)$.}

The occupation probabilities
\beq
f_{k,\eq}=\frac{p_k}{P(1)}\approx\frac{(b-1)\Gamma(b)}{k^b}
\label{afkc}
\eeq
fall off as a power-law in the thermodynamic limit.
The critical free energy reads
\be
\F_c=-\ln P(1)=-\ln\frac{b}{b-1}.
\ee
The second moment of the occupation probabilities,
\be
\mu_c=\sum_{k\ge0}k^2\,f_{k,\eq}=\frac{P'(1)+P''(1)}{P(1)}
=\frac{b+1}{(b-2)(b-3)},
\ee
is convergent for $b>3$ (regime of normal fluctuations),
and divergent for $2<b<3$ (regime of anomalous fluctuations).

\subsubsection*{Condensed phase ($\rho>\rho_c$).}

A large and finite system in the condensed phase
essentially consists of a uniform critical background,
containing on average $N_c=M\rho_c$ particles,
and of a macroscopic condensate, containing on average
$\D=N-N_c=M(\rho-\rho_c)$ excess particles with respect to the critical state.

The occupation probabilities $f_{k,\eq}$ accordingly split
into two main contributions~\cite{maj}.
The contribution of the critical background,
corresponding to small values of the occupation ($k\ll M)$,
is approximately given by~(\ref{afkc}).
The contribution of the condensate shows up as a hump located around $k=\D$.
The hump is a Gaussian whose width scales as $M^{1/2}$ whenever $\mu_c$
is finite, i.e., for $b>3$, whereas it has power-law tails and a larger width,
scaling as $M^{1/(b-1)}$, in the regime of anomalous fluctuations $(2<b<3)$.
The weight of the condensate probability hump is approximately $1/M$,
in accord with the picture that the system typically
contains a well-defined condensate located on a single site at any given time.

For the investigation of the stationary dynamics of the condensate,
we need to estimate the occupation probabilities
$f_{k,\eq}$ all over the probability dip region lying between the critical
background and the condensate hump, i.e., when $k\gg1$ and $\D-k\gg1$.
To reach this goal, we first estimate the partition function $Z_{M,N}$
{\it deep in the condensed phase}, i.e., for $\D=M(\rho-\rho_c)\gg1$.
An analogous analysis, with a different goal, is given in~\cite{evans2}.
In this regime, the contour integral in~(\ref{contour})
is dominated by the vicinity of the singular point $z=1$.
We therefore set $z=\e^{-\eps}$,
anticipating that the relevant $\eps$ will be small.
Equation~(\ref{contour}) then reads
\beq
Z_{M,N}=P(1)^M\int_\C\frac{\d\eps}{2\pi\i}
\exp\left[M\left((\rho-\rho_c)\eps+\cdots+A\,\eps^{b-1}+\cdots\right)\right].
\label{zmnc}
\eeq
In this expression, the contour $\C$ encircles the real negative axis,
whereas the first dots stand for higher-order regular terms
($\eps^2$, $\eps^3$, and so on)
and the second dots stand for higher-order singular terms.
The leading contribution comes from linearising
with respect to the leading singular term:
\be
Z_{M,N}\approx AM\,P(1)^M\int_\C\frac{\d\eps}{2\pi\i}\,\e^{\D\eps}\,\eps^{b-1}.
\ee
The contour integral reads $1/(\Gamma(1-b)\,\D^b)$.
We are thus left with
\beq
Z_{M,N}\approx(b-1)\Gamma(b)\frac{M}{\D^b}\left(\frac{b}{b-1}\right)^M.
\label{zas}
\eeq
(This expression also depends on $M$ and $N$ through $\D=N-M\rho_c$.)
The last exponential factor is $\exp(-M\F_c)$:
the free energy throughout the condensed phase is equal to the critical one.
Note the exact correspondence with the formula~(\ref{2sitb})
for the two-site system for $b>1$,
up to the replacement of $M$ by 2 and of $\D$ by $N$.
Equation~(\ref{afkeq}) then yields
\beq
f_{k,\eq}\approx(b-1)\Gamma(b)\frac{\D^b}{k^b(\D-k)^b}.
\label{afkdip}
\eeq
This estimate holds throughout the dip region ($k\gg1$ and $\D-k\gg1$).
The occupation probabilities~(\ref{afkdip}) are proportional
to the product $p_kp_{\D-k}$.
The normalisation matches with~(\ref{afkc}) for $1\ll k\ll\D$.
The above results are fully insensitive to the (normal or anomalous)
character of critical fluctuations.

The natural interpretation of the result~(\ref{afkdip})
is that typical configurations where one site contains $k$ particles
in the dip region ($k\gg1$ and $\D-k\gg1$)
are configurations where the remaining $\D-k$ excess particles
live on a single other site.
The dip region is therefore entirely dominated
by configurations where the excess particles are shared by {\it two} sites.
It can be checked that configurations where the excess particles
are shared by $p\ge3$ sites (i.e., $k+k_1+\cdots+k_{p-1}=\D$
with $k\gg1$ and all the $k_i\gg1$) have a negligible total weight.
This weight indeed comes from integrating the term $(A\eps^{b-1})^{p-1}$
in the right-hand-side of~(\ref{zmnc}).
It is therefore suppressed by a power-law factor $1/\D^{(p-2)(b-1)}$
with respect to the leading result~(\ref{afkdip}).

The derivation of~(\ref{afkdip})
can be extended to the ZRP with an arbitrary critical mass distribution.
The occupation probabilities in the dip region are still
dominated by configurations where the excess particles are shared by two sites.
They read
\beq
f_{k,\eq}\approx\frac{p_kp_{\D-k}}{p_\D P(1)}.
\label{gfkdip}
\eeq

\subsection*{Rates $u_k=1+a/k^\s$: Stretched-exponential critical behaviour}

We now turn to a brief account of the static properties of the ZRP with
hopping rate~\cite{wis1,wis2}
\beq
u_k=1+\frac{a}{k^\s},
\label{grates}
\eeq
where $\s$ is an arbitrary exponent.
The situation of interest corresponds to $0<\s<1$.
Equation~(\ref{pk}) leads to the estimate
\beq
p_k\sim\exp\left(-a\sum_{\ell=1}^k\frac{1}{\ell^\s}\right)
\sim\exp\left(-\frac{a}{1-\s}\,k^{1-\s}\right).
\label{gdecay}
\eeq
The generating series $P(z)$ has an essential singularity at $z=1$
with an exponentially small discontinuity.
The occupation probabilities at the critical density, $f_{k,\eq}=p_k/P(1)$,
decay as a stretched exponential law.

For a large finite system, the stationary occupation probabilities
in the dip region are given by~(\ref{gfkdip}), i.e., in the present situation,
\be
f_{k,\eq}\sim\exp\left(\frac{a}{1-\s}
\left[\D^{1-\s}-k^{1-\s}-(\D-k)^{1-\s}\right]\right).
\ee

\setcounter{equation}{0}
\def\theequation{B.\arabic{equation}}
\section*{Appendix~B.~Biased diffusion on an interval}

This Appendix gives a brief presentation of the properties of
a biased random walk on an interval in continuous time,
with arbitrary nearest-neighbour hopping rates.
The main emphasis is on the various definitions of characteristic times
and on their relationships.
Part of the content of this Appendix is classical
and can be found in the literature~\cite{std}.

\subsubsection*{Master equation.}

Sites are labeled by $k=0,\dots,N$.
The hopping rates are $\la_k$ to the right $(k\to k+1)$,
and $\mu_k$ to the left $(k\to k-1)$.
The boundaries are reflecting: $\la_N=\mu_0=0$.
The probability $f_k(t)$ that the random walk is at site $k$ at time $t$
obeys the master equation
\beqa\label{rat}
\frac{\d f_k}{\d t} &=&\mu_{k+1}f_{k+1}+\la_{k-1}f_{k-1}-(\mu
_k+\la_k)f_k\qquad(1\le k\le N-1),\nonumber\\
\frac{\d f_0}{\d t} &=&\mu_1f_1-\la_0f_0,\\
\frac{\d f_N}{\d t} &=&\la_{N-1}f_{N-1}-\mu_Nf_N.\nonumber
\eeqa
This equation can be recast as
\be
\frac{\d f_k}{\d t}=J_{k-1}-J_k,
\ee
by introducing the probability current through the bond $(k,k+1)$:
\be
J_k=\la_kf_k-\mu_{k+1}f_{k+1}.
\ee

\subsubsection*{Stationary state.}

In the stationary state, the current $J_{k,\eq}$ is independent of $k$.
The boundary conditions impose that it vanishes.
The detailed balance condition
\beq
\la_kf_{k,\eq}=\mu_{k+1}f_{k+1,\eq}
\label{deba}
\eeq
is therefore fulfilled, implying that the system is at equilibrium.
Equation~(\ref{deba}) yields
\beq
f_{k,\eq}=f_{0,\eq}\prod_{\ell=1}^k\frac{\la_{\ell-1}}{\mu_\ell},
\label{fkstb}
\eeq
where $f_{0,\eq}$ is fixed by normalisation.
The occupation probabilities can be rewritten as
\be
f_{k,\eq}=f_{0,\eq}\e^{-V_k},
\ee
in terms of the effective potential\footnote{In this paper we indifferently
use words describing the occupation probabilities themselves (hump, dip),
and the associated effective potential (valley, barrier).}
\be
V_k=\sum_{\ell=1}^k\ln\frac{\mu_\ell}{\la_{\ell-1}}.
\ee

\subsubsection*{Spectrum of characteristic times.}

The master equation~(\ref{rat}) reads formally
\be
\frac{\d f_k}{\d t}=\sum_\ell\M_{k,\ell}f_\ell,
\ee
where $\M$ is the Markov matrix of the process.
The spectrum of characteristic times is obtained by diagonalising $\M$.
Setting $f_k(t)=\varphi_k\,\e^{-E t}$, we obtain
\beq
\mu_{k+1}\varphi_{k+1,a}+\la_{k-1}\varphi_{k-1,a}-(\mu_k+\la_k)\varphi_{k,a}
=-E_a\,\varphi_{k,a}.
\label{spectr}
\eeq
The ground state $E_0=0$ corresponds to the stationary solution
$\varphi_{k,0}=f_{k,\eq}$ of~(\ref{fkstb}).
The characteristic times $\tau_a$ are the inverses
of the eigenvalues $E_a$ for $a=1,\dots,N$.

\subsubsection*{Mean first-passage time.}

Let $T_{k,0}$ be the mean first-passage time by the origin
of the particle with initial position~$k$,
i.e., the mean time it takes the particle to reach site~0.
This mean first-passage time obeys the following backward equation,
obtained by conditioning on the first step of the walker during
the infinitesimal duration~$\d t$,
\be
T_{k,0}=\d t+\left(1-(\la_k+\mu_k)\d t\right)T_{k,0}
+\la_k\d t\,T_{k+1,0}+\mu_k\d t\,T_{k-1,0},
\ee
hence
\beq
(\la_k+\mu_k)T_{k,0}-\la_kT_{k+1,0}-\mu_kT_{k-1,0}=1.
\label{back}
\eeq

The backward equation~(\ref{back}), with boundary condition $T_{0,0}=0$,
can be solved in closed form.
Indeed the differences $d_k=T_{k,0}-T_{k-1,0}$, defined for $k\ge1$,
obey the two-term equations $\mu_k d_k-\la_k d_{k+1}=1$.
The homogeneous equations, without the right-hand-side,
have the solution $d_k=1/(\mu_kf_{k,\eq})$.
Looking for a solution of the inhomogeneous equations in the form
$d_k=c_k/(\mu_kf_{k,\eq})$, we obtain the difference equation
$c_k-c_{k+1}=f_{k,\eq}$, with formally $c_{N+1}=0$, hence
$c_k=\sum_{\ell=k}^N f_{\ell,\eq}$, and finally
\beq
T_{k,0}=\sum_{\ell=1}^k\frac{1}{\mu_\ell
f_{\ell,\eq}}\sum_{m=\ell}^Nf_{m,\eq}.
\label{eq:fp}
\eeq

We now define the following two crossing times:
$T_\L\equiv T_{N,0}$ (resp.~$T_\R\equiv T_{0,N}$)
is the mean time it takes the particle to cross the system from right to left
(resp.~from left to right),
i.e., the mean time taken by the particle at initial position~$N$
to reach site~0 (resp.~at initial position~0 to reach site~$N$).
These crossing times read
\beq
T_\L=\sum_{\ell=1}^N\frac{1}{\mu_\ell f_{\ell,\eq}}
\sum_{m=\ell}^Nf_{m,\eq},\qquad
T_\R=\sum_{\ell=1}^N\frac{1}{\mu_\ell f_{\ell,\eq}}
\sum_{m=0}^{\ell-1}f_{m,\eq}.
\label{xtimes}
\eeq
The sum of these two times is the return crossing time $R$,
that is the mean time taken by the particle to cross the system
from one end of the system to the other and back:
\beq
T_\L+T_\R=R=\sum_{\ell=1}^N\frac{1}{\mu_\ell f_{\ell,\eq}}
=\sum_{\ell=0}^{N-1}\frac{1}{\la_\ell f_{\ell,\eq}}.
\label{sdef}
\eeq

The expressions derived so far are quite general.
They can be simplified in the following situations of interest.

\subsubsection*{The symmetric random walk.}

The case of the usual symmetric random walk
cor\-res\-ponds to uniform unbiased hopping rates
$\la_k=\mu_k=\la$ at all sites $k$.
Equation~(\ref{spectr}) becomes the discrete Laplace equation
\be
\la(\varphi_{k+1}+\varphi_{k-1}-2\varphi_k)=-E\varphi_k,
\ee
with boundary conditions $\varphi_{-1}=\varphi_0$ and
$\varphi_N=\varphi_{N+1}$.
The stationary state is uniform:
\be
f_{k,\eq}=\frac{1}{N+1}.
\ee
The other eigenvalues of the Markov matrix read
$E_a=4\la\sin^2 [a\pi/(2(N+1))]$.
Using~(\ref{eq:fp}), the mean first-passage by the origin reads
$T_{k,0}=k(2N+1-k)/(2\la)$,
hence the crossing times have the simple expression
$T_\L=T_\R=(N(N+1))/(2\la)$.
In the limit of a large interval $(N\gg1)$, we have therefore
\be
T_\L=T_\R\approx\frac{N^2}{2\la},\qquad\tau_1\approx\frac{N^2}{\pi^2\la},
\ee
hence
\beq
T_\L=T_\R\approx\frac{\pi^2}{2}\,\tau_1.
\label{trapp1}
\eeq
The crossing times and the main characteristic time grow
proportionally to the diffusive timescale $N^2$.

\subsubsection*{The case of a high potential barrier.}

The situation of most relevance in the context of this paper is when
the effective potential $V_k$ exhibits a high barrier,
i.e., a pronounced maximum $V_B\gg1$ at some position $k=B$.
The barrier defines two potential valleys, the left one ($0\le k<B$),
and the right one ($B<k\le N$).
For simplicity, we assume that the effective potential
is small everywhere except in the region of the barrier ($k\approx B$).

The respective stationary weights of the valleys read
\be
P_\L=\sum_{k=0}^{B-1}f_{k,\eq},\qquad P_\R=\sum_{k=B+1}^{N}f_{k,\eq}.
\ee
These two weights sum up approximately to unity
since $f_{B,\eq}\sim\e^{-V_B}$ is negligible.
The expressions~(\ref{xtimes}) for the crossing times simplify
in this regime.
Sums over $\ell$ are dominated by the contributions of sites $\ell\approx B$
close to the top of the potential barrier.
We thus obtain the estimates
\beq
T_\L\approx P_\R R,\qquad T_\R\approx P_\L R,
\label{tltr}
\eeq
since the return crossing time $R$ (see~(\ref{sdef})) is also
dominated by the vicinity of the top of the potential barrier.

The main characteristic time $\tau_1=1/E_1$ is also expected
to be very large in the presence of a high potential barrier.
It can be estimated as follows.
Let $\varphi_{k,1}$ be the corresponding eigenvector.
Introducing the ratios
\be
h_k=\frac{\varphi_{k,1}}{f_{k,\eq}},
\ee
Equations~(\ref{spectr}) are equivalent to
\beq
J_k-J_{k-1}=E_1 f_{k,\eq} h_k,\qquad
h_k-h_{k+1}=\frac{J_k}{\mu_{k+1}f_{k+1,\eq}}.
\label{ej}
\eeq
First, the current $J_k$ can be neglected, so that $h_k$ is constant.
This scheme holds as long as the $f_{k,\eq}$ are not too small,
i.e., separately in each valley.
We have therefore
\be
h_k\approx a_\L\quad(0\le k<B),\qquad h_k\approx a_\R\quad(B<k\le N).
\ee
The sum rule $\sum_{k=0}^N\varphi_{k,1}=0$ then implies
$a_\L\approx P_\R$, $a_\R\approx -P_\L$,
up to an overall multiplicative constant.
Second, the first equation in~(\ref{ej}) then yields the current,
to first order in $E_1$:
\be
J_k\approx P_\R E_1\sum_{\ell=0}^k f_{\ell,\eq}\quad(0\le k<B),\quad
J_k\approx P_\L E_1\sum_{\ell=k+1}^N f_{\ell,\eq}\quad(B<k\le N).
\ee
Both expressions consistently give $J_B\approx P_\R P_\L E_1$.
Third, using the second equation of~(\ref{ej}), we have
\be
1\approx P_\L+P_\R\approx a_\L-a_\R\approx h_0-h_N\approx J_B R
\approx P_\R P_\L E_1 R.
\ee
We thus finally obtain the following estimate for the main characteristic time:
\beq
\tau_1\approx P_\R P_\L R,
\label{t1}
\eeq
as well as the relationships
\be
\tau_1\approx P_\L T_\L\approx P_\R T_\R,
\qquad
\frac{1}{\tau_1}\approx\frac{1}{T_\L}+\frac{1}{T_\R}
\ee
between the main characteristic time and the crossing times.

If the potential barrier is very high, i.e., if $f_\min$ is very small,
a crude estimate of the characteristic times can be obtained by means of
a saddle-point evaluation of the sums in~(\ref{xtimes}):
\beq
\tau_1\sim T_\L\sim T_\R\sim\frac{1}{f_\min}\sim\e^{V_B}.
\label{arr}
\eeq
The Arrhenius law~(\ref{arrmin}) is thus recovered.
The corresponding activation energy is just the top
of the potential barrier, $V_B=-\ln f_\min$.

The full expressions~(\ref{tltr}) and~(\ref{t1}) however provide
more accurate predictions for the characteristic times in every situation
pertaining to the high-barrier regime.
They indeed lead to the correct absolute prefactor to the estimate~(\ref{arr}),
which usually involves a power of $N$.
For instance it is used in the present work in situations
where the characteristic times grows as a power of $N$.
The above mentioned prefactor is needed in general
in order to correctly predict this power.
Finally, let us mention the following self-consistent viewpoint on
the relevance of the high-barrier regime:
it is expected to hold whenever the predicted characteristic times
grow much faster than the diffusive timescale $N^2$.
This viewpoint is illustrated by the case of two sites (see section 4):
the high-barrier regime corresponds to $b>1$.

For the case of a symmetrically biased diffusion problem, the two valleys
are equally weighted ($P_\L=P_\R=1/2$).
The crossing time $T=T_\R=T_\L$ and the main characteristic time $\tau_1$
are thus given by
\beq
\tau_1\approx\frac{T}{2}=\frac{R}{4}.
\label{tsym}
\eeq
For a very asymmetric problem, with a small right valley of weight $P_R\ll1$
and a large left valley of weight $P_\L\approx1$,
the crossing time to the large valley
is close to the main characteristic time ($T_\L\approx\tau_1$),
whereas the crossing time to the small valley is much longer,
as it reads approximately $T_\R\approx\tau_1/P_\R$ in general.


\section*{References}

\begin{thebibliography}{99}

\bibitem{zeta1}
Bialas P, Burda Z and Johnston D 1997 Nucl. Phys. B {\bf 493} 505
\nonum
Bialas P, Burda Z and Johnston D 1999 Nucl. Phys. B {\bf 542} 413
\nonum
Bialas P, Bogacz L, Burda Z and Johnston D 2000 Nucl. Phys. B {\bf 575} 599

\bibitem{zeta2}
Drouffe J M, Godr\`eche C and Camia F 1998 J. Phys. A {\bf 31} L19

\bibitem{zeta3}
Godr\`eche C and Luck J M 2001 Eur. Phys. J. B {\bf 23} 473

\bibitem{barc}
Godr\`eche C and Luck J M 2002 J. Phys. Cond. Matt. {\bf 14} 1601

\bibitem{loan}
O'Loan O J, Evans M R and Cates M E 1998 Phys. Rev. E {\bf 58} 1404

\bibitem{evans1}
Evans M R 2000 Braz. J. Phys. {\bf 30} 42

\bibitem{cg}
Godr\`eche C 2003 J. Phys. A {\bf 36} 6313

\bibitem{gross}
Grosskinsky S, Sch\"utz G M and Spohn H 2003 J. Stat. Phys. {\bf 113} 389

\bibitem{evans2}
Evans M R and Hanney T 2005 J. Phys. A {\bf 38} R195

\bibitem{maj}
Majumdar S N, Evans M R and Zia R K P 2005 Phys. Rev. Lett. {\bf 94} 180601

\bibitem{wis1}
Kafri Y, Levine E, Mukamel D, Sch\"utz G M and T\"{o}r\"{o}k J 2002
Phys. Rev. Lett. {\bf 89} 035702

\bibitem{wis2}
Kafri Y, Levine E, Mukamel D, Schutz G M and Willmann R D 2003
Phys. Rev. E {\bf 68} R035101

\bibitem{pk}
Evans M R, Levine E, Mohanty P K and Mukamel D 2004 Eur. Phys. J. B {\bf 41} 223

\bibitem{glm}
Godr\`eche C, Levine E and Mukamel D 2005 J. Phys. A {\bf 38} L523

\bibitem{spitz}
Spitzer F 1970 Advances in Math. {\bf 5} 246

\bibitem{andj}
Andjel E D 1982 Ann. Prob. {\bf 10} 525

\bibitem{ar}
Arrhenius S 1889 Z. Phys. Chem. (Leipzig) {\bf 4} 226

\bibitem{kr}
Kramers H A 1940 Physica {\bf 7} 284

\bibitem{activ}
Godr\`eche C, Bouchaud J P and M\'ezard M 1995 J. Phys. A {\bf 28} L603
\nonum
Lipowski A 1997 J. Phys. A {\bf 30} L91
\nonum
Murthy K P N and Kehr K W 1997 J. Phys. A {\bf 30} 6671

\bibitem{ruin}
Feller W 1966 {\it An Introduction to Probability Theory and its Applications}
(New-York: Wiley) vol~1

\bibitem{kawa}
Cordery R, Sarker S and Tobochnik J 1981 Phys. Rev. B {\bf 24} 5402
\nonum
Cornell S J, Kaski K and Stinchcombe R B 1991 Phys. Rev. B {\bf 44} 12263
\nonum
Cornell S J and Bray A J 1996 Phys. Rev. E {\bf 54} 1153
\nonum
Ben-Naim E and Krapivsky P L 1998 J. Stat. Phys. {\bf 93} 583
\nonum
Godr\`eche C and Luck J M 2003 J. Phys. A {\bf 36} 9973

\bibitem{dgg}
Drouffe J M, Godr\`eche C and Grosskinsky S in preparation

\bibitem{std}
Risken H 1984 {\it The Fokker-Planck Equation: Methods of Solution and
Applications} (Berlin: Springer)
\nonum
van Kampen N G 1992 {\it Stochastic Processes in Physics and Chemistry}
(Amsterdam: North-Holland)
\nonum
Redner S 2001 {\it A Guide to First-Passage Processes} (Cambridge: CUP)
\nonum
Gardiner C W 2004 {\it Handbook of Stochastic Methods for Physics, Chemistry,
and Natural Sciences} Springer Series in Synergetics (Berlin: Springer)

\end{thebibliography}
\end{document}